\documentclass[epj]{webofc}
\usepackage[varg]{txfonts}   
\usepackage{graphicx}
\wocname{EPJ Web of Conferences}
\woctitle{ICNFP 2015}
\newcommand{\be}{\begin{equation}}
\newcommand{\ee}{\end{equation}}
\newcommand{\bea}{\begin{eqnarray}}
\newcommand{\eea}{\end{eqnarray}}

\newcommand{\rL}{\rho_{\Lambda}}

\begin{document}
\selectlanguage{english}

\begin{flushleft} 
\vspace*{0.5cm}
LCTS/2015-47,\,  KCL-PH-TH/2015-\textbf{57}
\end{flushleft} 

\title{Supersymmetry, Cosmological Constant and Inflation:\\
Towards a fundamental cosmic picture via ``running vacuum''}
%
%

\author{Nick E. Mavromatos\inst{1,2}\fnsep\thanks{\email{Nikolaos.Mavromatos@kcl.ac.uk}} 
}

\institute{Theoretical Particle Physics and Cosmology Group, Department of Physics, King's College London, Strand, London WC2R 2LS, UK
\and
          Theory Division, Physics Department, CERN CH 1211 Geneva 23, Switzerland
}

\abstract{%
  On the occasion of a century from the proposal of General relativity by Einstein, I attempt to tackle some open issues in modern cosmology, via a toy but non-trivial model. Specifically, I would like to link together:  (i) the smallness of the cosmological constant today, (ii) the evolution of the universe from an inflationary era after the big-bang till now, and (iii) local supersymmetry in the gravitational sector (supergravity) with a broken spectrum  at early eras, by making use of the concept of the ``running vacuum'' in the context of a simple toy model of four-dimensional N=1 supergravity. The model is characterised by dynamically broken local supersymmetry, induced by the formation of gravitino condensates in the early universe. As I will argue, there is a Starobinsky-type inflationary era characterising the broken supersymmetry phase in this model, which is compatible with the current cosmological data, provided a given constraint is satisfied among some tree-level parameters of the model and the renormalised cosmological constant of the de Sitter background used in the analysis. Applying the "running vacuum'' concept, then, to the effective field theory at the exit of inflation, makes a smooth connection (in cosmic time) with the radiation dominance epoch and subsequently with the current era of the Universe, characterised by a small (but dominant) cosmological-constant contribution to the cosmic energy density. In this approach, the smallness of the cosmological constant today is attributed to the failure (due to quantum gravity non-perturbative effects) of the aforementioned constraint.}
\maketitle

\section{Introduction and Summary} 

The year 2015 celebrates a century since the proposal by Einstein for the theory of General Relativity (GR), an elegant and quite successful (from a phenomenological point of view) classical field theory of Gravitation, generalising Newtonian gravity. 
GR pioneered the idea of linking the geometry of the cosmos with its dynamics, by attributing the curvature of space time to mass/energy of matter/radiation. The theory has two versions, \emph{local} - describing the dynamics of space-time in the neighbourhood of bodies, including celestial objects, and \emph{global}, dealing with the entire Universe as a whole, the latter being described today by the celebrated Friedmann-Lemaitre-Robertson-Walker (FLRW) model for Cosmology. The FLRW model describes quite successfully the plethora of the currently available data on Cosmology in the framework of the so-called $\Lambda$C(old)D(ark)M(atter) ($\Lambda$CDM) model.
Yet, despite these successes, there are a number of unanswered fundamental questions in both versions of GR, concerning: the quantisation of the theory, the microscopic nature of black holes and Hawking radiation, an explanation of the smallness of the (observed) cosmological constant today, 
a detailed microscopic theory of inflation and its graceful exit, as well as the nature of dark matter and dark energy of the Universe. 

On the other side of fundamental physics, particle theory, there are also great successes but also great mysteries which still remain open. The Standard Model (SM) of Electroweak and Strong interactions, is an elegant quantum field theory that unifies three of the four forces of nature quite successfully from an experimental point of view, but leaves gravity completely out of its reach. Moreover, there can be no explanation within the SM context of the matter-over-antimatter dominance in the universe today~\cite{bau}. In addition, the existence of dark matter and dark energy poses a pressing need for an extension of SM to accommodate the dark sector of the universe, \emph{e.g.} by incorporating supersymmetry (SUSY)~\cite{susy,akulov}, which could provide natural candidates for Dark Matter. 
Current searches in colliders at present, however, are not providing concrete evidence for such a framework, although it must be said that this may not be unnatural, given that the scale of supersymmetry breaking - depending on the SUSY model - may be out of reach of the current energies. 
Other non-trivial extensions of SM, include extra dimensional theories, as well as the whole string/brane framework, which might provide a consistent quantum framework for the unification of gravitational with the rest of the fundamental interactions in nature. In the latter framework, supersymmetry plays a crucial r\^ole for the stability of the vacuum in the majority of models, although there may be ways round it. Yet, no experimental evidence for the existence of strings or extra dimensions exists today. 

Motivated by the lack of concrete evidence for low energy extensions of the SM, but attributing an important r\^ole to the supersymmetrisation of the world, not so from the stability of the vacuum view point, but rather from the perspective that supersymmetry is the maximal space-time symmetry allowed by the Coleman-Mandula theorem~\cite{cm}, and thus worthy of further investigation, we would like to dedicate this talk to a study of a toy model of the Universe, which  could provide a non-trivial r\^ole of supersymmetry in ensuring inflation, graceful exit from it and subsequently connection with the FLRW cosmology until the current era, characterised by a small cosmological constant. 

The inflationary paradigm~\cite{inflation} offers an elegant solution to the horizon and flatness problems of the standard Big Bang cosmology, whilst simultaneously seeds both the large-scale structure of the universe and temperature anisotropies of the CMB via quantum fluctuations occurring during the inflationary epoch. Thus it seems a desirable concept to keep it in any successful cosmological model for the universe evolution. 
The precise microscopic mechanism of inflation is however unknown at present. The current data are in agreement with a scalar field (inflaton) (or fields) with canonical kinetic term(s) slowly rolling down an almost flat potential in the context of Einstein gravity, generating in the process 50 - 60 e-folds of inflation, along with adiabatic, nearly scale invariant primordial density perturbations \cite{Planck,encyclo}.
From the best fit value of the running spectral index $n_s\sim0.96$ for the gravitational perturbations in the slow-roll parametrisation, found by Planck \cite{Planck},  and the usual relations among the slow-roll inflationary parameters~\cite{encyclo}
$ n_s = 1-6\epsilon+2\eta\,, r = 16\epsilon$, 
one may deduce  $r\le 0.11$, given the non-observation of primordial gravitational wave-like (transverse and traceless) perturbations by Planck or WMAP collaborations (that is the absence of B-mode polarisations). This observational fact implies that the energy scale $E_I$ of inflation is much smaller than 
the Planck scale $m_P$, in particular  it lies in the ballpark of the scale of Grand Unification (GUT)~\cite{Planck,encyclo}:
$E_I \; = \; \Big(3\, H^2_I M_{\rm Pl}^2 \Big)^{1/4} \simeq 2.1 \times 10^{16} \times \Big(\frac{r}{0.20}\Big)^{1/4} \, {\rm GeV},$
with $M_{\rm Pl} = 2.4 \times 10^{18}~{\rm GeV}$  the reduced Planck mass, $H_I$ the Hubble scale during inflation and $r$ the tensor-to-scalar perturbation ratio~\cite{encyclo}. The upper bound on $r < 0.11$ placed by the Planck Collaboration~\cite{Planck} implies 
\begin{equation}\label{upperH}
H_I  \leq 0.81 \times 10^{14}~{\rm GeV}\, \leqslant 3.38 \times10^{-5} \, M_{\rm Pl}~,
\end{equation}
which is five orders of magnitudes smaller than the reduced Planck mass. 
This hierarchy between inflation scale and Planck scale, already prompted researchers to seek for a natural reason for it, and already in the early days after the inflation paradigm was at play, links of the low scale of inflation with supersymmetry have been made~\cite{natural}.

If supersymmetry is realised in nature however, it is certainly broken, with partners masses above TeV scale, given that they have not been discovered at the large hadron collider at present. Simple realisations of global supersymmetry (SUSY) breaking, such as in the Wess-Zumino model~\cite{croon}, can provide, 
when embedded in gravitational environments, slow-roll models for inflation consistent with the Planck data~\cite{Planck}.
Rigorous embeddings of global SUSY to local supersymmetry (SUGRA) have also been considered and explored in the literature over the years in connection with various scenarios for inflation~\cite{sugra_inflation_review}, such as hybrid~\cite{sugra_hybrid}, chaotic~\cite{sugra_chaotic}, no-scale SUGRA~\cite{sugra_staro}, Jordan-frame-SUGRA~\cite{confsugra} and others~\cite{sugrainfl}.  In the case of \emph{some} no-scale SUGRA models, the inflationary potential can be made similar to that of the so-called Starobinsky model of inflation~\cite{staro}, by a specific choice of parameters. Unlike the SUGRA case, however, in the original Starobinsky model, the inflaton 
scalar mode represents collectively the effects of the quadratic-order scalar curvature terms ($R^2$) in the effective gravitational action, obtained after integrating out (conformal) matter in the early hot Universe~\cite{staro}.
Starobinsky-type models seem to be preferred by the Planck data~\cite{Planck}.

In this talk I shall review a rather minimal inflationary scenario which is associated indirectly with a Starobinsky type inflation in its original sense~\cite{staro}. 
Specifically, I will discuss the appearance of scalar inflationary modes 
associated with higher curvature terms of the (one-loop) effective action of simple (3+1)-dimensional 
$\mathcal{N}=1$ SUGRA models without matter~\cite{Freedman,Nieuwenhuizen}, obtained by integrating out massive gravitino degrees of freedom in the broken supergavity phase. The supersymmetry breaking occurs dynamically via the formation of gravitino condensates, as a consequence of the four-gravitino interactions that characterise (any) supergravity action, via the fermionic torsion parts of the spin connection. 
The approach is documented in a series of previous publications~\cite{emdyno,ahm,ahmstaro} and will be reviewed briefly here. We should stress that our approach does not supersymmetrise higher curvature terms in the action to obtain Starobinksy-like effective actions, as \emph{e.g.} in the case of \cite{terada}. For us, the higher curvature terms emerge, as we mentioned, as a result of quantum loop corrections in the broken-local-supersymmetry phase of the $\mathcal N =1$ simple SUGRA action, after path-integrating out massive gravitino fields and graviton fluctuations in a de Sitter background (where such integration becomes meaningful~\cite{fradkin}).
 The dynamical breaking process may be concretely realised by means of a phase transition from the supersymmetric phase where the bilinear $\langle\overline\psi_\mu\psi^\mu\rangle$ representing the effective scalar degree of freedom has zero vacuum expectation value, to one where $\sigma \equiv \langle\overline\psi_\mu\psi^\mu\rangle\neq0$. 
The quantum excitations about this condensate vacuum are then identified with a gravitino condensate scalar field. 
Since this must be an energetically favourable process to occur, it then follows that the effective potential experienced by the gravitino condensate must be locally concave about the origin. The corresponding one-loop effective potential of the gravitino condensate scalar field, therefore, has the characteristic form of a Coleman-Weinberg double well potential, offering the possibility of hilltop-type inflation, with the condensate field playing the role of the inflaton~\cite{emdyno,ahm}, which would be the simplest scenario. However,  
in order to guarantee a slow-roll inflation one needs unnaturally large values of the gravitino-condensate wave function renormalisation, unless transplanckian scales for supersymmetry breaking are invoked.
This prompted us to discuss a second~\cite{ahmstaro}, rather indirect way, by means of which the gravitino condensate field is associated with inflation. This is realised via 
the higher (in particular quadratic-order) curvature corrections of the (quantum) effective action of the gravitino condensate field, obtained after path-integration of graviton and gravitino degrees of freedom in the massive gravitino phase. These corrections induce a 
Starobinsky-type inflation~\cite{staro}, which occurs for quite natural values of the parameters of the ${\mathcal N}=1$ SUGRA model (actually, its Jordan-frame variants~\cite{confsugra}), with the inflation scale and gravitino mass in the ball park of the GUT scale.

The structure of the talk is as follows: In Section \ref{sec:effpot} we review the formalism of the dynamical breaking of SUGRA in 
four dimensional ${\mathcal N} =1$ models, including superconformal extensions thereof (with broken conformal symmetry) that are necessitated for phenomenological reasons, as we shall see. In Section \ref{sec:star} we discuss scenarios for inflation of Starobinsky type that may occur in the massive gravitino phase, near the non-trivial minimum of the effective potential. In such scenarios, which are compatible with the Planck satellite data~\cite{Planck}, the r\^ole of the inflaton field is played by the scalar mode that describes collectively the effects of scalar-curvature-squared terms that characterise the gravitational sector of the one-loop effective action in the broken SUGRA phase, after integrating out the massive gravitinos. In section \ref{sec:rvm} we apply~\cite{basil} the running vacuum concept~\cite{rvm} in the effective field theory at the exit of the Starobinsky inflation, in order to ensure a smooth (in cosmic time) connection of that era with radiation- and matter-dominance eras of the Universe, and subsequently with the current era characterised by a small but dominant cosmological-constant type contribution to the cosmic energy density. 
Finally, conclusions and outlook are presented in section \ref{sec:concl}.

\section{Dynamical breaking of ${\mathcal N}=1$ $D=4$ SUGRA  models \label{sec:effpot}} 

This section is based on material presented in refs.~\cite{ahm}. We consider the (on-shell) action for `minimal' $\mathcal{N}=1$ $D=4$ supergravity in the second order formalism~\cite{Freedman,Nieuwenhuizen}:
 \be\label{sugraction} S_{\rm{SG}}=\int d^4x \,e \left(\frac{1}{2\kappa^2}R\left(e\right)-\overline\psi_\mu\gamma^{\mu\nu\rho}D_\nu\psi_\rho+\mathcal{L}_{\rm torsion}\right), 
	\, \gamma^{\mu\nu\rho}=\frac{1}{2}\left\{\gamma^\mu,\gamma^{\nu\rho}\right\}\,,
	\, \gamma^{\nu\rho}=\frac{1}{2}\left[\gamma^\nu,\gamma^\rho\right]\,,
\ee
where  $\kappa^2 = 8\pi {\rm G} = 1/M_{\rm Pl}^2$, in units $\hbar=c=1$, $M_{\rm Pl}$ is the reduced Planck mass in four space-time dimensions, 
$e$ is the vierbein determinant and the scalar space-time curvature $R(e)$ and gravitational covariant derivative $D_\nu\psi_\rho\equiv\partial_\nu\psi_\rho+\frac{1}{4}\omega_{\nu ab}\left(e\right)\gamma^{ab}\psi_\rho$ are defined via the torsion-free connection~\footnote{Our metric signature is $(-, +, +, +)$ and the definitions of the Ricci
and Riemann curvature tensors are  $R_{\mu\nu} = R^\lambda_{\mu \lambda \nu}$  and  $R^\lambda_{\mu\nu\rho} = \partial_\nu \, \Gamma_{\mu\rho}^\lambda - \dots $, respectively, with $\Gamma_{\mu\nu}^\lambda  = \Gamma_{\nu\mu}^\lambda $ the torsion-free Christofel symbol, and the Ricci scalar is given by $R = g^{\mu\nu} R_{\mu\nu}$.}. In the gauge condition 
\begin{equation} \gamma_\mu\, \psi^\mu =0~,
\label{gcond}
\end{equation}
we shall work with in this article, 
the  torsion term $\mathcal{L}_{\rm torsion}$, arising  from the fermionic torsion parts of the spin connection, reads: 
\be\label{torsion}
	\mathcal{L}_{\rm torsion}=-\frac{1}{8} \, \kappa^2\,\left(\left(\overline\psi^\rho\gamma^\mu\psi^\nu\right)\left(\overline\psi_\rho\gamma_\mu\psi_\nu+2\overline\psi_\rho\gamma_\nu\psi_\mu\right)\right).
\ee
We notice at this stage that the Fierz identities among the gravitinos, make the actual coefficients of these four-fermion terms ambiguous. This is a known ambiguity in the context of mean field theory, and  only the non-perturbative physics can settle the value of the four-fermion terms~\cite{Wetterich}. Thus we may extend the definition of the torsion four-fermion interaction terms to include such ambiguities and write:
\be\label{4ftorsion}
		\mathcal{L}_{\rm torsion}
		=\lambda_{\rm S}\left(\overline\psi^\rho\psi_\rho\right)^2
		+\lambda_{\rm PS}\left(\overline\psi^\rho\gamma^5\psi_\rho\right)^2
		+\lambda_{\rm PV}\left(\overline\psi^\rho\gamma^5\gamma_\mu\psi_\rho\right)^2
\ee
where the couplings $\lambda_{\rm S}$, $\lambda_{\rm PS}$ and $\lambda_{\rm PV}$ express the freedom we have to rewrite each quadrilinear in terms of the others via Fierz transformation. They satisfy the constraint: 
$\lambda_{\rm S} - \lambda_{\rm PS} + 4\lambda_{\rm PV} = -3\kappa^2/4$.
This freedom was addressed in the second reference in \cite{ahm}, where we refer the reader for details.

Following the original ideas of dynamical symmetry breaking by Nambu and Jona-Lasinio~\cite{NJL}, we wish to linearise these four-fermion interactions via suitable auxiliary fields, \emph{e.g.}
$$	\frac{1}{4}\left(\overline\psi^\rho\psi_\rho\right)^2\sim\sigma\left(\overline\psi^\rho\psi_\rho\right)-\sigma^2,$$
where the equivalence (at the level of the action) follows as a consequence of the subsequent Euler-Lagrange equation for the auxiliary scalar $\sigma$.
Our task is then to look for a non-zero vacuum expectation value $\langle\sigma\rangle$ which would serve as an effective mass for the gravitino. 
To induce the super-Higgs effect~\cite{DeserZumino} we also couple in the Goldstino, associated to global supersymmetry breaking, via~\cite{akulov}
\be\label{goldstino}
	\mathcal{L}_\lambda=f^2\det\left(\delta_{\mu\nu}+\frac{i}{2f^2}\overline\lambda\gamma_\mu\partial_\nu\lambda\right)\bigg|_{\gamma\cdot\psi=0}=f^2+\dots\,,
\ee	
where $\lambda$ is the Goldstino, $\sqrt{f}$ expresses the scale of global supersymmetry breaking, and \dots\, represents higher order terms which may be neglected in our weak-field expansion of the determinant.
It is worth emphasising at this point the universality of (\ref{goldstino}); any model containing a Goldstino may be related to $\mathcal{L}_\lambda$ via a non-linear transformation \cite{Komargodski}, and thus the generality of our approach is preserved. 

Upon the specific gauge choice (\ref{gcond}) for the gravitino field
and an appropriate redefinition, one may eliminate any presence of the Goldstino field from the final effective 
action describing the dynamical breaking of local supersymmetry, except the cosmological constant term $f^2$ in (\ref{goldstino}), which serves as a reminder of the pertinent scale of supersymmetry breaking.
The non-trivial energy scale this introduces, along with the disappearance (through field redefinitions) of the Goldstino field from the physical spectrum and the concomitant development of a gravitino mass, characterises the super-Higgs effect~\cite{DeserZumino}.
We may then identify in the broken phase an effective action
\be\label{finalaction}
	S=\frac{1}{2\kappa^2}\int d^4x \,e \left(\left(R\left(e\right)-2\Lambda\right)-\overline\psi_\mu\gamma^{\mu\nu\rho}D_\nu\psi_\rho+ m_{\rm dyn} \, \left(\overline\psi_\mu \psi^\mu\right)\right),	
\ee
where $\Lambda$ is the renormalised cosmological constant, to be contrasted with the (negative) tree level cosmological constant 
\be\label{l0}
	 \Lambda_0\equiv\kappa^2\left(\sigma ^2  -f^2\right)~,
\ee
and $m_{\rm dyn} \propto \langle \sigma \rangle $ is a dynamically generated gravitino mass, the origin of which will be explained presently.
It is worth stressing at this point that $\Lambda_0$ must be negative due to the incompatibility of supergravity with de Sitter vacua; if SUGRA is broken at tree level, then of course no further dynamical breaking may take place.

For phenomenological reasons which have been outlined in detail in refs.~\cite{emdyno,ahm}, and we shall discuss below, we adopt an extension of ${\mathcal N}=1$ SUGRA which incorporates local supersymmetry in the Jordan frame, enabled by an associated dilaton superfield~\cite{confsugra}. 
The scalar component $\varphi$ of the latter can be either a fundamental space-time scalar mode of the gravitational multiplet, i.e. the trace of the graviton (as happens, for instance, in supergravity models that appear in the low-energy limit of string theories), or a composite scalar field constructed out of matter multiplets.
In the latter case these could include the standard model fields and their superpartners that characterise the Next-to-Minimal Supersymmetric Standard Model, which can be consistently incorporated in such Jordan frame extensions of SUGRA~\cite{confsugra,nmssm}, leading to supersymmetric generalisations of the Higgs inflation~\cite{higgsinfl}. 
Upon appropriate breaking of conformal symmetry, induced by specific dilaton potentials (which we do not discuss here), one may assume that the dilaton field acquires a non-trivial vacuum expectation value $\langle \varphi \rangle \ne 0 $. 
One consequence of this is then that in the broken conformal symmetry phase, the resulting supergravity sector, upon passing (via appropriate field redefinitions) to the Einstein frame is described by an action of the form (\ref{sugraction}), but with the coupling of the gravitino four-fermion interaction terms being replaced by 
\begin{equation}\label{tildcoupl}
\tilde \kappa\equiv e^{-\langle\varphi\rangle}\kappa~,
\end{equation}
while the Einstein term in the action carries the standard gravitational coupling $1/2\kappa^2$. In this way the v.e.v. of the dilaton may be combined with the Fierz-induced 
ambiguities (\ref{4ftorsion}), so that the non-perturbative physics that determines the latter also determines the scale of the conformal symmetry breaking. As in our approach we do not discuss details of the dilaton potential, we shall consider from now on such a v.e.v., as well the coefficient $\lambda_S$ of the  four fravitino terms in (\ref{4ftorsion}) leading to the formation of scalar gravitino condensates,  of interest to us here, as phenomenological. 
\begin{figure}[h!!]
		\centering
		\includegraphics[width=0.7\textwidth]{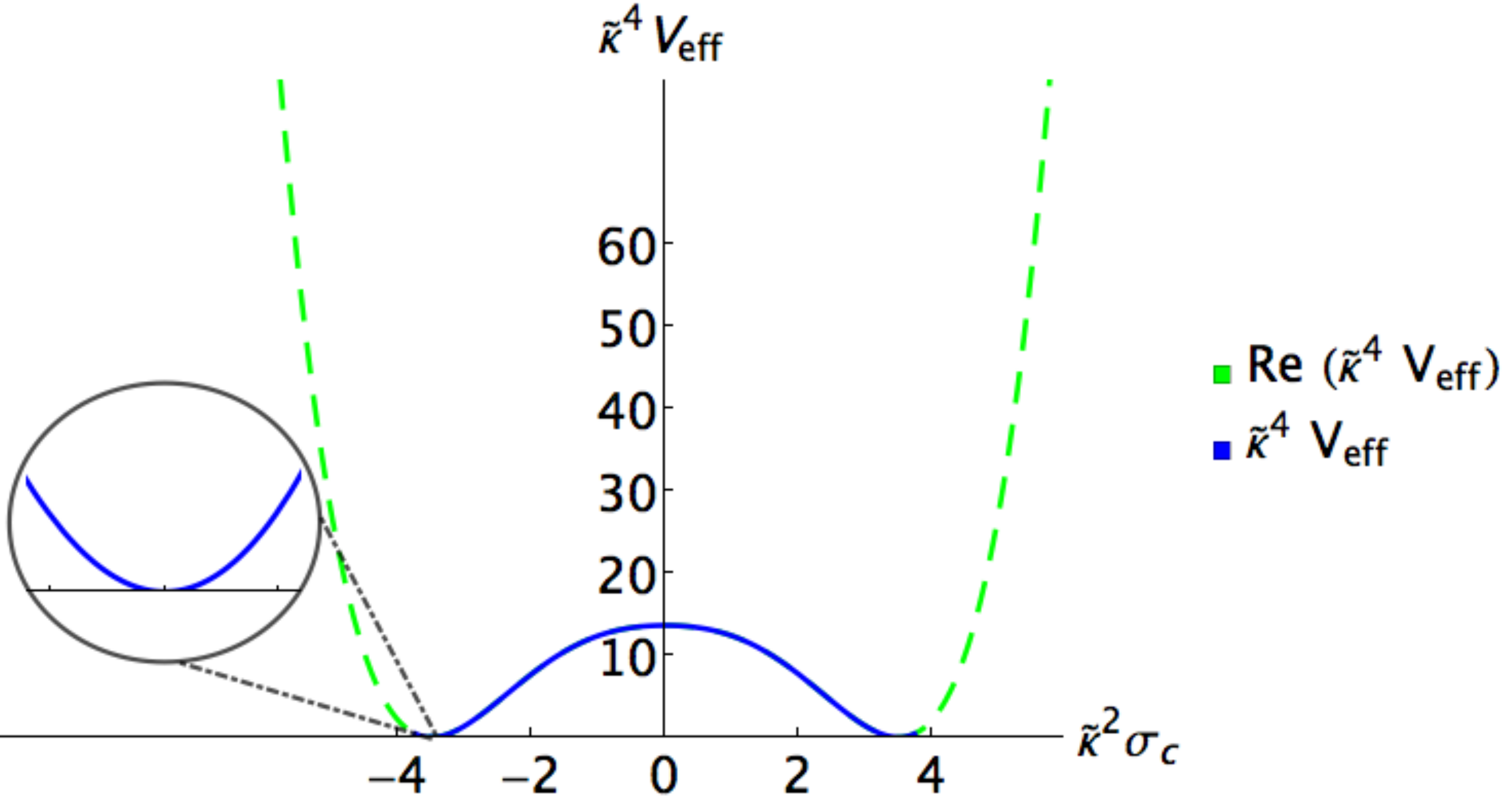} \vfill
		\includegraphics[width=0.4\textwidth]{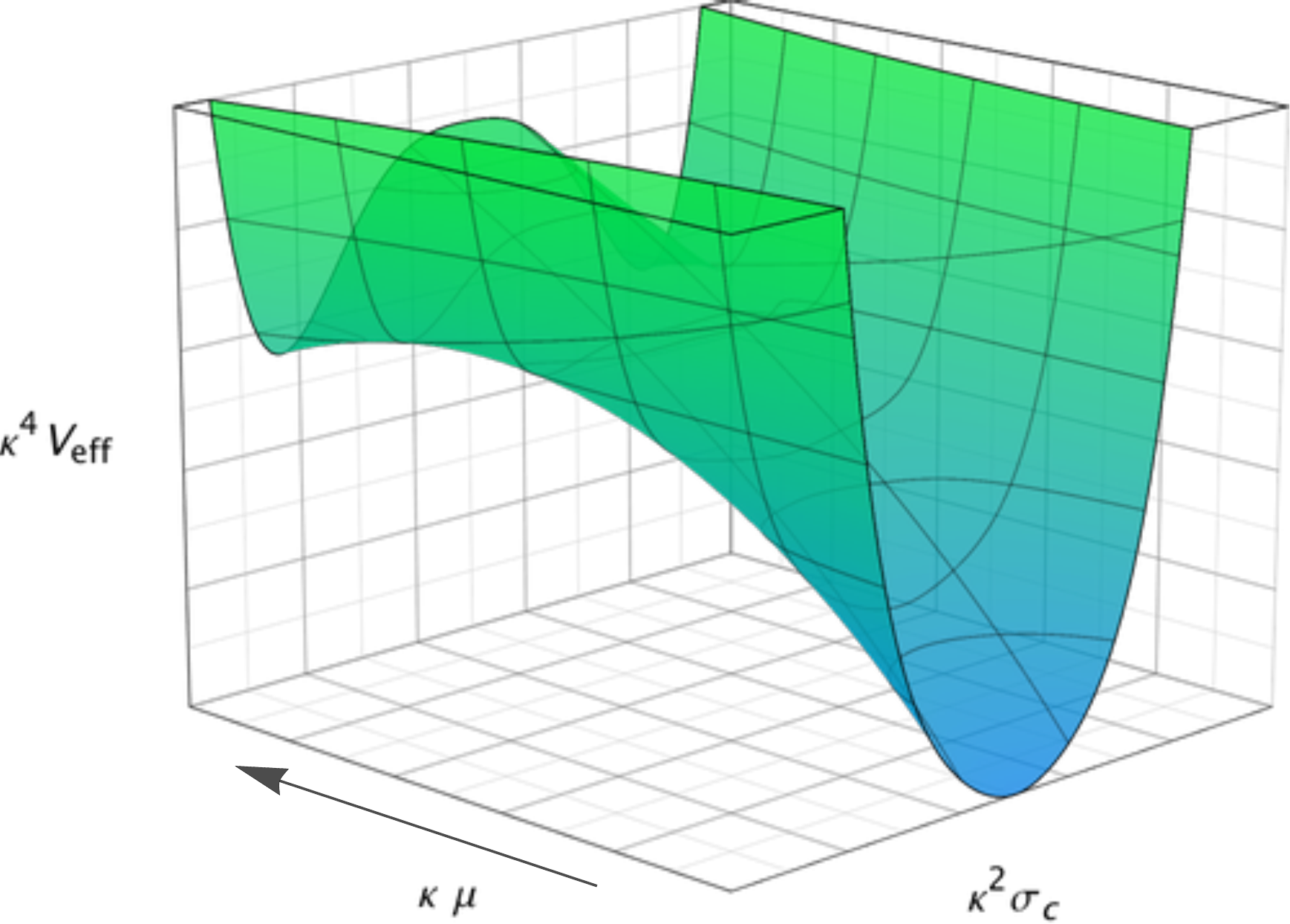}
		\includegraphics[width=0.4\textwidth]{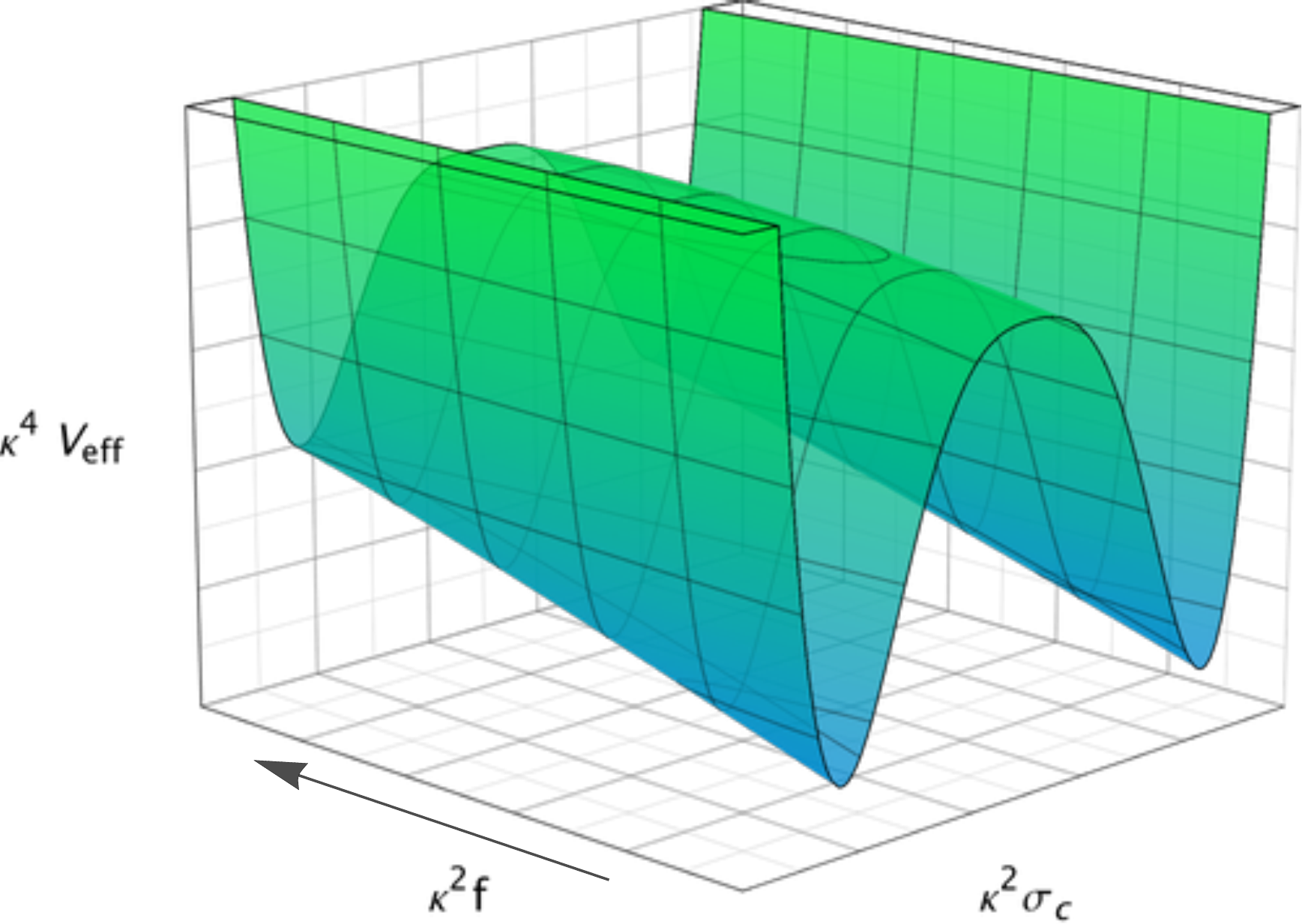}
		\caption{\emph{Upper panel}: The one-loop effective potential (\ref{effpotsugra}), expressed in units of the 
		coupling $\tilde \kappa$ (\ref{tildcoupl}). \emph{Lower panel}: As above, but showing schematically the effect of tuning the  RG scale $\mu$ and the supersymmetry breaking scale $f$, whilst holding, respectively, $f$ and $\mu$ fixed. The arrows in the respective axes correspond to the direction of increasing $\mu$ and $f$. Pictures taken from ref.~\cite{ahm}.}
\label{fig:effpotsugra}	
	\end{figure}
Expanding the graviton field about a de Sitter background~\cite{fradkin} (under the assumption that it is a solution of the one-loop effective equations) with renormalised cosmological constant $\Lambda > 0$, and integrating out both bosonic and fermionic quantum fluctuations to one loop yields the following effective potential for the gravitino condensate field $\sigma$ in the flat space-time limit $\Lambda \to  0$, as detailed in ref.~\cite{ahm}, 
\be\label{effpotsugra}
		V_{\rm{eff}}=V_{B}^{(0)}+V_{B}^{(1)}+V_{F}^{(1)}
		=-\frac{\Lambda_0}{\kappa^2}+V_{B}^{(1)}+V_{F}^{(1)}~,
	\ee
where $\Lambda_0$ is given by (\ref{l0}) and 
	\begin{eqnarray}\label{boson}
	&	V_{B}^{(1)}=\frac{45 \kappa ^4}{512 \pi^2}\left(f^2-\sigma ^2\right)^2 \left(3-2 \ln \left(\frac{3 \kappa ^2 \left(f^2-\sigma ^2\right)}{2 \mu ^2}\right)\right)~, \nonumber \\
		&V_{F}^{(1)} = \left(\frac{{\tilde \kappa}}{\kappa}\right)^4 \, \frac{\kappa^4 \sigma^4}{30976 \pi ^2} \left(30578 \ln \left(\left(\frac{{\tilde \kappa}}{\kappa}\right)^2\, \frac{\kappa^2 \sigma^2}{3 \mu ^2}\right)-45867+29282 \ln \left(\frac{33}{2}\right)+1296 \ln\left(\frac{54}{11}\right)\right)~,
	\end{eqnarray}
indicate the contributions to the effective potential from bosonic and fermionic fields respectively, and 	
$\mu$ is an inverse renormalisation group (RG) scale, 
associated with a proper time short distance cut-off~\cite{ahm},  such that the flow from the ultraviolet to the infrared limit corresponds to the direction of an increasing $\mu$. 
The one-loop effective potential (\ref{effpotsugra}) is depicted in fig.~\ref{fig:effpotsugra} and it has the characteristic double-well shape of dynamical breaking of a symmetry. We note that as we flow from UV to IR (\emph{i.e}. in the direction of increasing $\mu$),  we obtain the correct double-well shape required for the super-Higgs effect, and secondly that tuning $f$ allows us to shift $V_{\rm{eff}}$ and thus attain the correct vacuum structure (\emph{i.e.} non-trivial minima $\sigma_c$ such that $V_{\rm{eff}}\left(\sigma_c\right)=0$).
Moreover,  the shape of the effective potential changes, as one varies the 
(renormalisation) scale $\mu$ from ultraviolet to infrared values (\emph{i.e.} flowing in the direction of increasing $\mu$), in such a way that the broken symmetry phase (double-well shaped potential) is reached in the IR. 
This indicates that the dynamical generation of a gravitino mass is actually an IR phenomenon, in accordance with the rather general features of dynamical mass in field theory.  In the broken phase, the mass of the gravitino condensate is then given by 
$m_\sigma^2\equiv V''_{\rm{eff}}(\sigma_c),$
where $\sigma_c$ is the minimum of $V_{\rm{eff}}$ and a prime denotes a functional derivative with respect to the gravitino-condensate field. As observed from (\ref{boson}), the bosonic contributions to the effective potential contain logarithmic terms which would contribute imaginary terms, leading to instabilities, unless 
\be\label{imagin}
\sigma_c^2<f^2~.
\ee
From (\ref{effpotsugra}) it is straightforward to see that this condition is equivalent to the negativity of the tree-level cosmological constant $\Lambda_0$, which is entirely sensible; if $\Lambda_0>0$ then SUGRA is broken at tree level (given the incompatibility of supersymmetry with de Sitter vacua) and there can be no dynamical breaking.
As such, we must then tune $f$ for a given value of $\mu$ to find self consistent minima $\sigma_c$ satisfying (\ref{imagin}), thereby ensuring a real $V_{\rm{eff}}$. In fact, here lies the importance of the super-Higgs effect, and thus of a non-zero positive $f^2 > \sigma_c^2 > 0$, in allowing dynamical breaking of local supersymmetry~\footnote{In refs.~\cite{odintsov}, the importance of the super-Higgs effect was ignored, which led to the incorrect conclusion that imaginary parts exist necessarily in the one-loop effective potential, in the same class of gauges as in ref.~\cite{ahm} and here, and hence dynamical breaking of SUGRA was not possible. As we have seen above, such imaginary parts are absent when the condition (\ref{imagin}) is satisfied, and thus dynamical breaking of SUGRA occurs.}. 

As discussed in refs.~\cite{emdyno,ahm}, phenomenologically realistic situations, where one avoids transplanckian gravitino masses, for supersymmetry breaking scales $\sqrt{f} $ at most of order of the Grand Unification (GUT) scale $10^{15-16}$~GeV, as expected from arguments related to the stability of the electroweak vacuum, can occur only for 
large $\tilde \kappa$ couplings, typically of order $\tilde \kappa \sim \Big(10^{3}-10^{4} \Big)\, \kappa$. 
Given the relation (\ref{tildcoupl}) this corresponds to dilaton vev of $\mathcal{O}\left(-10\right)$, where the negative sign may be familiar in the context of dilaton-influenced cosmological scenarios \cite{aben}.

If we consider for concreteness the case $\tilde\kappa=10^3 \kappa$, which is a value dictated by the inflationary phenomenology of the model~\cite{emdyno}, we may find solutions with a vanishing one-loop effective potential at the non-trivial minima corresponding to:
${\tilde \kappa}^2 \, \sigma_c \simeq 3.5~, \quad {\tilde \kappa}^2 \, f \simeq 3.7~, \quad {\tilde \kappa}\, \mu \simeq 4.0,$
which leads to a global supersymmetry breaking scale 
	\be\label{fscaleconf}
		\sqrt{f} \simeq 4.7\times10^{15}~{\rm GeV}~,
	\ee
 and dynamical gravitino mass 
 	\be\label{gravinoconf}
 		m_{\rm dyn}  
		\simeq 2.0\times10^{16}~{\rm GeV}~.
	\ee
	
At the non-trivial minima we find $\tilde\kappa^4V_{F}^{(1)}\simeq-1.4$, $\tilde\kappa^4V_{B}^{(1)}\simeq5.9\times10^{-13}$, with tree-level cosmological constant $\tilde\kappa^2\Lambda_0 \simeq-1.4$.
We thus observe that fermion contributions to the effective potential are much stronger than the corresponding bosonic contributions for the cases of large couplings $\tilde \kappa \gg \kappa$. 
These values are phenomenologically realistic, thereby pointing towards the viability (from the point of view of producing realistic results of relevance to phenomenology) of the scenarios of dynamical breaking of local supersymmetry in conformal supergravity models~~\footnote{A comment concerning SUGRA models in the Jordan frame with such large values for their frame functions is in order here. In our approach, the dilaton $\varphi$ could be a genuine (dimensionless) dilation scalar field $\varphi = 2\phi$ arising in the gravitational multiplet of strings, whose low-energy limit may be identified with some form of SUGRA action. In our normalization the string coupling would be $g_s \equiv e^\phi = {\tilde \kappa}^{-1/2}$. In 
such a case, a value of $\tilde \kappa = e^{-\langle \varphi \rangle}\, \kappa = {\mathcal O}(10^{3-4})$ would imply a large negative v.e.v. of the (four-dimensional) dilaton field of order $\langle \phi \rangle = -{\mathcal O}(5) < 0$, and thus a weak string coupling squared $g_s = {\mathcal O}(10^{-2})$, which may not be far from values attained in phenomenologically realistic string cosmologies~\cite{aben}. On the other hand, in the Jordan-frame SUGRA models of \cite{confsugra}, the frame function reads $\Phi \equiv e^{-\varphi} =  1 - \frac{1}{3}\Big(S{\overline S} + \sum_{u,d} \, H_i H^\dagger_i \Big) - \frac{1}{2} \chi \, \Big(-H_u^0 \, H_d^0 + H_u^+ \, H_d^- + {\rm h.c.} \Big)$,
in the notation of \cite{nmssm} for the various matter super fields of the next-to-minimal supersymmetric standard model that can be embedded in 
such supergravities. The quantity $\chi$ is a constant parameter. At energy scales much lower than GUT, it is expected that the various fields take on subplanckian values, in which case the frame function is almost one, and hence $\tilde \kappa \simeq \kappa $ for such models today. To ensure $\tilde \kappa \gg \kappa$,  and thus large values of the frame function, $\Phi \gg 1$, as required in our analysis, one needs to invoke trasnplanckian values for some of the fields, $H_{u,d}^0$, and large values of $\chi$, which may indeed characterize the inflationary phase of such theories. A similar situation occurs for the values of the higgs field (playing the role of the inflaton) in the non-supersymmetric Higgs inflation models~\cite{higgsinfl}.}. On the other hand, in standard SUGRA scenarios, where $\tilde \kappa = \kappa$, one finds, as already mentioned, transplanckian values for the dynamically generated gravitino mass~\cite{ahm}:
$m_{\rm dyn} \simeq 2.0\times10^{19}~{\rm GeV}$, and a 
global supersymmetry breaking scale $\sqrt{f} \simeq 4.7\times10^{18}~{\rm GeV}$, far too high
to make phenomenological sense.

\section{Starobinsky-type inflation in the broken SUGRA phase \label{sec:star}}

Starobinsky inflation is a model for obtaining a de Sitter (inflationary) cosmological solution to gravitational equations arising from a (four space-time-dimensional) action that includes higher curvature terms. 
Specifically, an action of the type in which the quadratic curvature corrections consist only of scalar curvature terms~\cite{staro}
\be\label{staroaction}
{\mathcal S} = \frac{1}{2 \, \kappa^2 } \, \int d^4 x \sqrt{-g}\,  \left(R  + \beta  \, R^2 \right) ~, \quad
\beta = \frac{8\, \pi}{3\, {\mathcal M}^2 }~,
\ee
where ${\mathcal M}$ is a constant of mass dimension one, characteristic of the model, which cannot be determined by theoretical considerations, and hence should be considered phenomenological. 

The important feature of this model is that inflationary dynamics is driven purely by the gravitational sector, through the $R^2$ terms, 
and that the scale of inflation is linked to ${\mathcal M}$. From a microscopic point of view, the scalar curvature-squared terms in (\ref{staroaction}) are viewed as the result of quantum fluctuations (at one-loop level)  of conformal (massless or high energy) matter fields of various spins, which have been integrated out in the relevant path integral in a curved background space-time~\cite{loop}. 

The above considerations necessitate truncation to one-loop quantum order and to curvature-square (four-derivative) terms, which 
implies that there must be a region of validity for curvature invariants such that $\mathcal{O}\big(R^2/M_{\rm Pl}^4\big) \ll 1$. 
This is of course a condition satisfied in phenomenologically realistic scenarios of inflation~\cite{Planck,encyclo}, for which the inflationary Hubble scale $H_I $ satisfies (\ref{upperH}) (the reader should recall that $R = 12 H_I^2$ in the inflationary phase). 

Although the inflation in this model is not driven by rolling scalar fields, nevertheless the model (\ref{staroaction}) (and for that matter, any other model where the Einstein-Hilbert space-time Lagrangian density is replaced by an arbitrary function $f(R)$ of the scalar curvature) is conformally equivalent to that of an ordinary Einstein-gravity coupled to a scalar field with a potential that drives inflation~\cite{whitt}. This can be seen 
by first linearising the $R^2$ terms in (\ref{staroaction}) by means of an auxiliary (Lagrange-multiplier) field $\tilde \alpha (x)$,  and then rescaling the metric by a conformal transformation and redefining the scalar field, so that the final theory acquires canonically-normalised Einstein and scalar-field terms:
\begin{eqnarray}\label{confmetric}
&&g_{\mu\nu} \rightarrow g^E_{\mu\nu} = \left(1 + 2 \, \beta \, {\tilde \alpha (x)} \right) \, g_{\mu\nu} ~, \quad
 \tilde \alpha \left(x\right) \to \rho (x) \equiv \sqrt{\frac{3}{2}} \, {\rm ln} \, \left(1 + 2\, \beta \, {\tilde \alpha \left(x\right)} \right)~.
\end{eqnarray}
These steps may be understood schematically via
\be\label{steps}
	\int d^4 x \sqrt{-g}\,  \left( R  + \beta  \, R^2 \right) 
   \hookrightarrow\int d^4 x \sqrt{-g^E}\,  \left(R^E +  g^{E\, \mu\, \nu} \, \partial_\mu \, \rho \, \partial_\nu \, \rho - V(\rho) \right)~,
\ee
where the arrows have the meaning that the corresponding actions appear in the appropriate path integrals, 
with the potential $V(\rho)$ given by:
\be\label{staropotent}
 V(\rho ) = \frac{\left( 1 - e^{-\sqrt{\frac{2}{3}} \, \rho } \right)^2}{4\, \beta} \, 
  = \frac{3 {\mathcal M}^2 \, \Big( 1 - e^{-\sqrt{\frac{2}{3}} \, \rho } \Big)^2}{32\, \pi }  \,  ~.
\ee
This potential is sufficiently flat for large values of $\rho$ (compared to the Planck scale) to produce phenomenologically acceptable inflation, with the scalar field $\rho$ playing the role of the inflaton. 
In fact, the Starobinsky model, with its characteristically low value of the tensor-to-scalar ratio $r$, provides an excellent fit to the recent Planck data on inflation~\cite{Planck}.

\begin{figure}[h!!!]
\centering
		\includegraphics[width=0.7\textwidth]{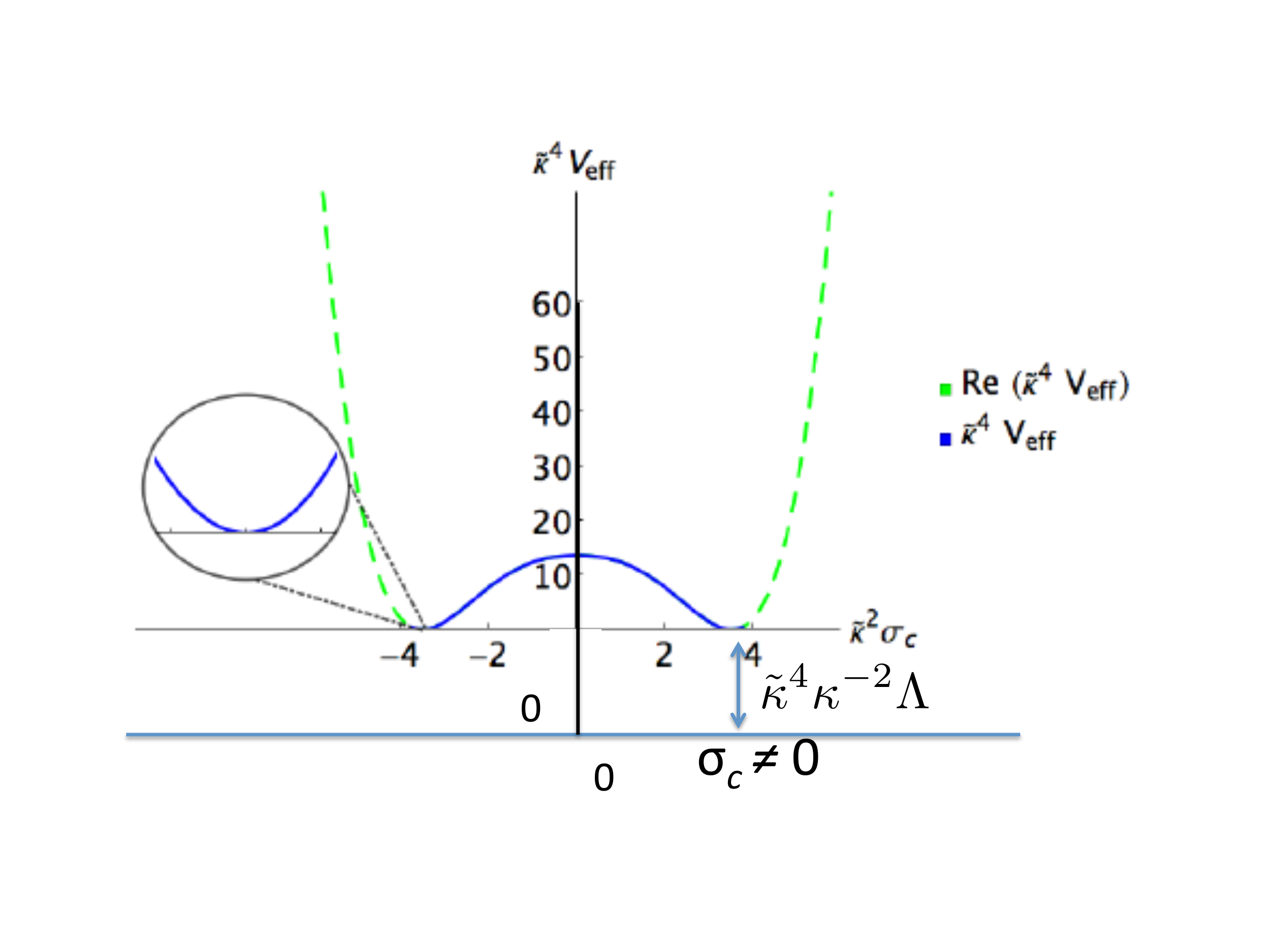}
		\caption{Generic shape of the one-loop effective potential for the gravitino condensate field $\sigma_c$ in dynamically broken (conformal) Supergravity models in the presence of a non-trivial de Sitter background with cosmological constant $\Lambda > 0$~\cite{ahmstaro}. The Starobinsky inflationary phase is associated with fluctuations of the condensate and gravitational field modes near the non-trivial minimum of the potential, where the condensate $\sigma_c \ne 0$, and the potential assumes the value $\Lambda > 0$, consistent with supersymmetry breaking. The dashed green lines denote ``forbidden'' areas of the condensate field values, violating the condition (\ref{imagin}), for which imaginary parts appear in the effective potential, thereby destabilising the broken symmetry phase.}
\label{fig:pot2}	
\end{figure}

Below we consider an extension~\cite{ahmstaro} of the dynamically broken SUGRA analysis of the previous section
 to the case where the de Sitter parameter $\Lambda$ is perturbatively small compared to $m_P^2$, but not zero, so that truncation of the series to order $\Lambda^2$ suffices. 
This is in the spirit of the original Starobinsky model~\cite{staro}, with the r\^ole of matter fulfilled by the now-massive gravitino field.
Specifically, we are interested in the behaviour of the effective potential near the non-trivial minimum, where $\sigma_c $ is a non-zero constant. 
It is important to notice at this point that, in contrast to the original Starobinsky model~\cite{staro}, where the crucial for inflation $R^2$ terms 
have been argued to arise from  the \emph{conformal anomaly} in the path integral of massless (conformal) matter in a de Sitter background, 
and thus their coefficient was arbitrary, in our scenario, such terms arise in the one-loop effective action of the gravitino condensate field, evaluated in a de Sitter background, after integrating out massive gravitino fields, whose mass was generated dynamically. The order of the de Sitter cosmological constant,
$\Lambda > 0$ that breaks supersymmetry, and 
the gravitino mass are all evaluated dynamically (self-consistently) in our approach from the minimization of the effective potential. Thus, the resulting $R^2$ coefficient, which determines the phenomenology of the inflationary phase,  is calculable~\cite{ahmstaro}. 
Moreover, in our analysis, unlike Starobinsky's original work, we will keep the contributions from both graviton (spin-two) and gravitino quantum fluctuations.
Specifically, we are interested in the behaviour of the effective potential near the non-trivial minimum, where $\sigma_c $ is a non-zero constant (\emph{cf.} fig. \ref{fig:pot2}). 

The one-loop effective potential, obtained by integrating out gravitons and (massive) gravitino fields in the scalar channel (after appropriate euclideanisation), may be expressed as a power series in the renormalised cosmological constant $\Lambda$~\cite{ahmstaro}: 
\be\label{effactionl2}
	\Gamma\simeq S_{\rm cl}-\frac{24\pi^2}{\Lambda^2 }\big(\alpha^F_0+\alpha_0^B
	+ \left(\alpha^F_{1}+ \alpha^B_{1}\right)\Lambda
	+\left(\alpha^F_{2}+ \alpha^B_{2}\right)\Lambda^2+\dots\big)~,
\ee 
where $S_{\rm cl} =  -\frac{1}{2\kappa^2}\int d^4 x \sqrt{g}\left(\widehat{R}-2\Lambda_0\right)$
is the classical action with $\Lambda_0$ given by (\ref{l0}) and $\widehat R$ denoting the fixed $S^4$ background we expand around ($\widehat R=4\Lambda$, Volume = $24\pi^2/\Lambda^2$); the $\alpha^{B(F)}_i$, $i=1,2$ in (\ref{effactionl2}) indicate the bosonic (B) and fermionic (F) quantum corrections at a given order ($i=1, 2$) in $\Lambda$, respectively.
The leading order term in $\Lambda$ is then the effective action found in \cite{ahm} in the limit $\Lambda\to0$, 
\bea		
&&\Gamma_{\Lambda\to0}\simeq-\frac{24\pi^2}{\Lambda^2}\left(-\frac{\Lambda_0}{\kappa^2}+\alpha_0^F+\alpha_0^B\right) \equiv\frac{24\pi^2}{\Lambda^2}\frac{\Lambda_1}{\kappa^2}, \nonumber \\	
&& \alpha_0^{F} = {\tilde \kappa}^4 \, \sigma_c^4 \, \Big(0.100\,  \ln \left( \frac{{\tilde \kappa}^2 \, \sigma_c^2}{3 \mu ^2}\right) + 0.126 \Big)~, \nonumber \\
&&	\alpha_0^{B}= \kappa ^4 \, \left(f^2-\sigma_c^2\right)^2 \left(0.027 - 0.018 \ln \left(\frac{3 \kappa ^2 \left(f^2-\sigma_c^2\right)}{2 \mu ^2}\right)\right)~.		
\eea
The remaining quantum corrections , proportional to $\Lambda$ and $\Lambda^2$, may be identified respectively with Einstein-Hilbert $R$-type and Starobinsky $R^2$-type terms in an effective action (\ref{effactionl3}) of the form
\be\label{effactionl3}
\Gamma\simeq -\frac{1}{2\kappa^2} \int d^4 x \sqrt{g} \left(\left(\widehat R-2\Lambda_1\right)  +\alpha_1 \, \widehat R+ \alpha_2 \, \widehat R^2\right)~,
\ee
where we have combined terms of order $\Lambda^2$ into curvature scalar square terms. For general backgrounds such terms 
would correspond to invariants of the form ${\widehat R}_{\mu\nu\rho\sigma} \, {\widehat R}^{\mu\nu\rho\sigma} $, ${\widehat R}_{\mu\nu} \, {\widehat R}^{\mu\nu}$ and ${\widehat R}^2$, which for a de Sitter background all combine to yield ${\widehat R}^2$ terms~\footnote{In general, however, the metric tensor fluctuations of such terms may differ from the ones due to the $R^2$ term alone. We note that, in four-dimensions, in the absence of non-trivial (\emph{i.e.} non-constant) dilatons, one may actually deal with just one more independent invariant, $R_{\mu\nu} R^{\mu\nu}$, in addition to $R^2$, since the Gauss-Bonnet combination, $R_{\mu\nu\alpha\beta}\, R^{\mu\nu\alpha\beta} - 4\, R_{\mu\nu} R^{\mu\nu} + R^2,$ is a total space-time derivative. This is not the case for Jordan-frame SUGRA, where non-trivial-dilaton factors accompany the higher-order curvature terms. Nevertheless, in the minimum of the dilaton potential, where the conformal symmetry is broken, the dilaton equals its v.e.v. which is a space-time constant and, hence, this case is equivalent to  the constant-dilaton one.}. 
The coefficients $\alpha_1$  and $\alpha_2$ absorb the non-polynomial (logarithmic) in $\Lambda$ contributions, so that we may then identify (\ref{effactionl3}) with (\ref{effactionl2}) via 
	\be\label{alpha}
		\alpha_1=\frac{\kappa^2}{2}\left(\alpha^F_1+\alpha^B_1\right)~,\quad
		\alpha_2=\frac{\kappa^2}{8}\left(\alpha^F_2+\alpha^B_2\right)~.
	\ee

To identify the conditions for phenomenologically acceptable Starobinsky inflation around the non-trivial minima of the broken SUGRA phase 
of our model, we impose first the \emph{cancellation} of the ``classical'' Einstein-Hilbert space term $\widehat R $ by the ``cosmological constant'' term $\Lambda_1$, \emph{i.e}. that 
\be\label{condition}
\widehat R = 4 \, \Lambda = 2\, \Lambda_1 ~.
\ee
This condition should be understood as a necessary one characterising our background in order to produce phenomenologically-acceptable 
Starobinsky inflation in the broken SUGRA phase following the first inflationary stage, as discussed in ref.~\cite{emdyno}. 
This may naturally be understood as a generalisation of the relation $\widehat R=2\Lambda_1=0$, imposed in ref.~\cite{ahm} as a self-consistency condition for the dynamical generation of a gravitino mass in a Minkowski space-time, discussed in the previous section. From (\ref{condition}) it follows that the cosmological constant $\Lambda$ satisfies the four-dimensional Einstein equations in the non-trivial minimum, and in fact coincides with the value of the one-loop effective potential of the gravitino condensate at this minimum. As we discussed in \cite{ahm}, this non-vanishing positive value of the effective potential is consistent with the generic features of dynamical breaking of supersymmetry~\cite{witten}. In terms of the Starobinsky inflationary potential
(\ref{staropotent}), the value $\Lambda > 0$ corresponds to the approximately constant value of this potential in the high $\varphi$-field regime, where Starobinsky-type inflation takes place. Thus we may set 
$\Lambda \sim 3\, H_I^2 $, where $H_I$ the (approximately) constant Hubble scale during inflation, which is constrained by the current data to satisfy (\ref{upperH}).

The effective Newton's constant in  (\ref{effactionl3}) is then 
\be\label{egc}
\kappa_{\rm eff}^2=\kappa^2/\alpha_1~, 
\ee
and from this, we can express the effective Starobinsky scale (\ref{staroaction}) in terms of $\kappa_{\rm eff}$ as $\beta_{\rm eff} \equiv  \alpha_2/\alpha_1$.
This condition thus makes a direct link between the action (\ref{effactionl2}) with a Starobinsky type action (\ref{staroaction}).
Comparing with (\ref{staroaction}), we can then identify the Starobinsky inflationary scale in this case as
\begin{equation}\label{staroours}
{\mathcal M} = \sqrt{\frac{8 \pi}{3} \, \frac{\alpha_1}{\alpha_2} }~.
\end{equation}

The coefficients $\alpha_1$ and $\alpha_2$ (\ref{alpha}) have been determined in ref.~\cite{ahm}:
	\begin{eqnarray}\label{aif}
\alpha^F_1&=& 0.067\, \tilde\kappa^2 \sigma_c ^2  -0.021\, \tilde\kappa^2 \sigma_c ^2 \, {\rm ln} \left(\frac{\Lambda}{\mu^2}\right) 
 + 0.073\, \tilde\kappa^2 \sigma_c ^2 \, {\rm ln} \left(\frac{\tilde\kappa^2\sigma_c^2}{\mu^2} \right)~, 
\nonumber \\
\alpha^F_{2}&=& 0.029 + 0.014\, {\rm ln} \left(\frac{\tilde\kappa^2\sigma_c^2}{\mu^2}\right)  -0.029\, {\rm ln} \left(\frac{\Lambda}{\mu^2}\right)~, \nonumber \\
\alpha^B_1&=& -0.083 \Lambda_0 + 0.018\, \Lambda_0 \, {\rm ln} \left(\frac{\Lambda }{3 \mu ^2}\right)  + 0.049\, \Lambda_0\,  {\rm ln} \left(-\frac{3 \Lambda_0}{\mu ^2}\right)~, 
\nonumber \\
\alpha^B_{2}&=& 0.020 +  0.021\, {\rm ln} \left(\frac{\Lambda }{3 \mu ^2}\right) - 0.014\, {\rm ln} \left(-\frac{6 \Lambda_0}{\mu ^2}\right)~,
\end{eqnarray}	
where $\sigma_c$ denotes the gravitino scalar condensate 
at the non-trivial minimum of the one-loop effective potential (\emph{cf.} fig.~\ref{fig:pot2}),
and 
${\tilde \kappa} = e^{-\langle \varphi  \rangle } \, \kappa$ is the conformally-rescaled gravitational constant in the model of ref.~\cite{confsugra}, defined previously via (\ref{tildcoupl}). 

In ref.~\cite{ahmstaro} we searched numerically for points in the parameter space such that  the effective equations
$\frac{\partial\Gamma}{\partial\Lambda}=0~, \quad
		 \frac{\partial\Gamma}{\partial\sigma}=0$, are satisfied; $\Lambda$ is small and positive ($0<\Lambda<10^{-5}M^2_{\rm Pl}$, to ensure the validity of our expansion in $\Lambda$) and $10^{-6}<\mathcal{M}/M_{\rm Pl}<10^{-4}$, to match with known phenomenology of  Starobinsky inflation \cite{Planck}. For $\tilde \kappa=\kappa$ (\emph{i.e}. for non-conformal supergravity), we were unable to find any solutions satisfying these constraints, which was to be expected, given the previously demonstrated non-phenomenological suitability of this simple model~\cite{ahm}. 
If we consider $\tilde \kappa>>\kappa$, however, for instance of order $\tilde \kappa / \kappa = {\mathcal O}(10^{3} - 10^4))$,
we find that we are able to satisfy the above constraints for a range of values~\cite{ahmstaro}, with
gravitino mass and global SUSY breaking scale in the ball-park of GUT scale, and the Starobinsky scale of order ${\mathcal M} \sim 10^{-5} \, M_{\rm Pl} $, which leads to phenomenologically acceptable inflation in the massive gravitino phase, consistent with the Planck-satellite data~\cite{Planck}. 
In particular,  typical values obtained for the parameters of our conformal SUGRA models satisfy~\cite{ahm,basil}
  \begin{equation}\label{scales}
  \Lambda \sim 3H_I^2 \sim m_{3/2}^2 \sim {\tilde \kappa}^2 \sigma_c^2 \sim \kappa^2 f^2 \ll \mu^2 = 8\pi/\kappa^2\,, \sigma_c^2 \ll f^2.
  \end{equation} 
 Since the scale of SUSY breaking must be in the ballpark of the typical GUT scale associated to the inflation, namely $\sqrt{f}\sim 10^{16}$ GeV$\sim 10^{-2}\,M_{Pl}$, from (\ref{scales}) we obtain  $\Lambda \sim \kappa^2f^2=f^2/M_{Pl}^2\sim 10^{27}$ GeV$^2$. As a result, the scale of the gravitino is some two to three orders of magnitude below the GUT scale, $m_{3/2}\sim \sqrt{\Lambda}\gtrsim 10^{13}$ GeV$\sim 10^{-5}\,M_{Pl}$, which 
is compatible with the bound (\ref{upperH}).

\section{``Running Vacuum'' and Broken SUGRA effective field theory \label{sec:rvm}}

Exit from the inflationary phase is a complicated issue which we shall not discuss here, aside from the observation that it can be achieved by
coherent oscillations of the gravitino condensate field around its minima. 
This is still an open issue, which may be addressed via construction of more detailed supersymmetric models, including coupling of the matter sector to gravity, which will determine the pattern of the inflaton decays. 

\subsection{The ``Running Vacuum'' Scenario and post inflationary evolution of the Universe \label{sec:evol}}

Nevertheless, non-trivial conclusions about the evolution of the Universe after the phase transition that describes graceful exit from inflation and a smooth connection (in the sense of an evolution in cosmic time) of this phase with the radiation-dominated and subsequent eras of the Universe till the current epoch, can be made by applying the concept of the ``running vacuum''~\cite{rvm} (RV) immediately after the exit from inflation~\cite{basil}.
According to the RV hypothesis~\cite{rvm}, the dynamical cosmic evolution (``decay'') of the inflationary phase ground state (``vacuum'') to the standard radiation regime can be described using a ``renormalization-group-like'' approach, whereby the time evolution of the vacuum energy density $\rho_\Lambda (t)$  is inherited from its dependence on a characteristic
cosmic scale variable $\mu_c=\mu_c(t)$. This variable plays the role
of running (mass) scale of the renormalization group (RG) approach, and a
natural candidate for such scale in FLRW cosmology is the Hubble
parameter $H(t)$. Therefore the proposed RG equation is
~\cite{rvm}:
\begin{equation}\label{runningrho}
\frac{d\, \rho_\Lambda (t)}{d\, {\rm ln}H^2 } = \frac{1}{(4\pi)^2}
\sum_i \Big[a_i M_i^2 H^2 + b_i H^4 + c_i \frac{H^6}{M_i^2} + \dots
\Big]
\end{equation}
The coefficients $a_i, b_i, c_i \dots$ appearing in
(\ref{runningrho}) are dimensionless and receive contributions from
loop corrections of boson and fermion matter fields with different
masses $M_i$. It must be stressed that the requirement of general covariance of the
action~\cite{rvm} implies 
\emph{only even} powers of the (cosmic-time $t$ dependent) Hubble
parameter $H(t)$ on the right-hand-side of (\ref{runningrho}).

We  note that, if the evolution of the Universe is
restricted to eras below the GUT scale, as is the case of the dynamically broken SUGRA discussed here, then
at most  the $H^4$ terms  in (\ref{runningrho}) can
contribute significantly to the $\rho_\Lambda (t)$ evolution.
Then, on Integrating the RG equation (\ref{runningrho}) one obtains: 
\bea\label{lambda}
&&\rL(H) = \frac{\tilde \Lambda(H)}{\kappa^2}=\frac{3}{\kappa^2}\left(c_0 + \nu H^{2} + \alpha
\frac{H^{4}}{H_{I}^{2}}\right) \;, \nonumber \\
&&\nu=\frac{1}{48\pi^2}\, \sum_{i=F,B} a_i\frac{M_i^2}{M_{\rm
Pl}^2}\,,  \quad \alpha = \frac{1}{96\pi^2}\, \frac{H_I^2}{M_{\rm Pl}^2}\sum_{i=F,B} b_i~.\eea
Here $c_0$ is an integration constant (with dimension $+2$ in
natural units, \emph{i.e.} energy squared) which can be fixed from the low energy data of the
current universe, and contributes to the cosmological constant.

The RV model is based on the assumption that at any moment in cosmic time $t$, 
the vacuum is characterised by the Equation of State (EoS) of de-Sitter space time,
\emph{i.e.} 
$ p_\Lambda (t) = - \rho_\Lambda (t) = \tilde \Lambda (t) / \kappa^2,$
where $\tilde \Lambda (t) $ denotes the vacuum energy of the RV model. 
We stress that this EoS does not depend on whether the vacuum is dynamical or
not. In contrast to other forms of dark energy, the vacuum is
defined as that for which the EoS parameter $\omega$ is precisely
$\omega=-1$ at any moment of the Universe's evolution.

Einstein's equations in this framework can be written as:
\begin{equation}
R_{\mu \nu }-\frac{1}{2}g_{\mu \nu }R=\kappa^2\,  \tilde{T}_{\mu\nu}\,,
\label{EE}
\end{equation}
where the total stress tensor $\tilde{T}_{\mu\nu}$ is given by
$\tilde{T}_{\mu\nu}\equiv T_{\mu\nu}-g_{\mu\nu}\,\rho_{\Lambda} $,
with $T_{\mu\nu}=-2\partial{\cal L}_{m}/\partial
g_{\mu\nu}+g_{\mu\nu}\,{\cal L}_{m}$ the energy-momentum tensor
corresponding to the  matter Lagrangian. The extra piece proportional to 
$\rho_{\Lambda}$, corresponds to the vacuum energy
density associated with the presence of $\tilde \Lambda(t)$ (with pressure
$p_{\Lambda}=-\rho_{\Lambda}$).
Modeling the expanding universe in this framework as a perfect fluid with velocity
$4$-vector field $U_{\mu}$, we have for the matter-radiation  stress tensor 
$T_{\mu\nu}=p_{m}\,g_{\mu\nu}+(\rho_{m}+p_{m})\,U_{\mu}U_{\nu}$,
where $\rho_{m}$ is the density of matter-radiation and
$p_{m}=\omega_{m} \rho_{m}$ is the corresponding pressure, in which
$\omega_m$ is the EoS of matter.  $\tilde{T}_{\mu\nu}$
takes the same form as ${T}_{\mu\nu}$ but with $\rho_{\rm
tot}=\rho_{m}+\rho_{\Lambda}$ and $p_{\rm
tot}=p_{m}+p_{\Lambda}=p_{m}-\rho_{\Lambda}$, that is,
$\tilde{T}_{\mu\nu}=
(p_{m}-\rho_{\Lambda})\,g_{\mu\nu}+(\rho_{m}+p_{m})U_{\mu}U_{\nu}$.

The integrated form (\ref{lambda}) of the vacuum energy density and the above considerations, 
provides an effective description of inflation and the subsequent stages of the FLRW Universe evolution in a smooth way~\cite{rvm,basil}.
In particular, as discussed in \cite{basil}, from Einstein's equations in the context of RV, 
and in particular the conservation equation (Bianchi identity) of the total stress tensor $\bigtriangledown^{\mu}\,{\tilde{T}}_{\mu\nu}=0$, 
one can deduce 
\begin{equation}
\dot{\rho}_{m}+3(1+\omega_{m})H\rho_{m}=-\dot{\rho_{\Lambda}}\,.
\label{frie33}
\end{equation}
and the evolution equation of the Hubble parameter $H(t)$:
\begin{equation}
\label{HE} \dot
H+\frac{3}{2}(1+\omega)H^2\left[1-\nu-\frac{c_0}{H^2}-\alpha\frac{H^2}{H_I^2}\right]=0\,,
\end{equation}
in units of the inflation scale $H_I$ (\emph{cf.} (\ref{upperH})). Eq.~(\ref{HE}) can be solved at various epochs~\cite{rvm,basil}. For instance, during inflationary era, one may neglect the term proportional to $c_0$ on the right-hand-side of 
(\ref{HE}), to find a constant $H$ as a self-consistent solution during inflation, 
$H^2=(1-\nu)H_I^2/\alpha$. Going away from the inflationary regime, the Hubble parameter is represented by a solution of the form
\begin{equation}\label{HS1}
 H(a)=\left(\frac{1-\nu}{\alpha}\right)^{1/2}\,\frac{H_{I}}{\sqrt{D\,a^{3(1-\nu)(1+\omega)}+1}}\,,
\end{equation}
where $D$ is a positive constant of integration and $a=a(t)$ is the scale factor of the universe. 
The universe will enter the standard radiation phase, with $w=1/3$, in the case when $Da^{4(1-\nu)} \gg 1$, which is confirmed by 
substituting this solution (\ref{HS1}) into (\ref{lambda}) to obtain:
\begin{equation}\label{eq:rLa}
  \rho_\Lambda(a)=\frac{{\rho}_I}{\alpha}\,\frac{1}{\left[1+D\,a^{4}\right]^{2}}\,,
 \end{equation} 
 where $\rho_I=3H_I^2/\kappa^2$ is the critical density in the inflationary epoch.
Then solving (\ref{frie33}) we obtain for the radiation energy density $\rho_m$:
\begin{equation}\label{eq:rhor}
 \rho_r(a)=\frac{{\rho}_I}{\alpha}\,\frac{D\,a^{4}}{\left[1+D\,a^{4}\right]^{2}}\,.
\end{equation}
From the above expressions, it is apparent that there is no singularity in the initial state: the Universe starts at $a=0$ with a huge vacuum energy density $\rho_I/\alpha$ (and zero radiation) which is progressively converted into relativistic matter. In the asymptotic radiation regime we indeed retrieve the standard behavior $\rho_r \sim  a^{-4}$ with essentially negligible vacuum energy density: $\rho_\Lambda\sim a^{-8}\ll\rho_r$. Graceful exit from inflation is, therefore, naturally implemented in this formalism. 

Subsequently, the universe will enter a  matter dominated phase, and then the current era, in which the term $c_0$ in (\ref{lambda}) becomes dominant.
In this case Eq.(\ref{lambda}) amounts to
\be\label{vcc}
\tilde \Lambda(H)= 3c_{0}+3\nu H^{2}_{0} 
+3\nu(H^{2}-H_{0}^{2}),
\ee which  
plays the r\^ole of the cosmological
constant at the present time~\cite{basil}, which is positive. The reader should notice the RV corrections 
to  the cosmic-concordance $\Lambda$CDM model. 

\subsection{Applying the ``Running Vacuum'' scenario  to the Dynamically Broken SUGRA model \label{sec:vmsugra}}

Let us now apply the above reasoning to the effective potential of the broken $\mathcal N = 1$ SUGRA model, at the exit of its inflationary phase~\cite{basil}. 
The effective (dynamical) vacuum energy density, $\rho_\Lambda(H)$,
during the inflationary phase of our SUGRA model can be extracted from the SUGRA effective action $\Gamma$ \,(\ref{effactionl3}),
upon applying the constraint (\ref{condition}) and analytically continuing the results back to Minkowski space-time signature.
In particular, the effective potential is defined as $V_{\rm eff} \equiv - \Gamma \rightarrow \int d^4 x \sqrt{-g}\, \rho_\Lambda (H)$.
Moreover, at the exit from inflation, any logarithmic dependence of the coefficients $\alpha_i$ on $\Lambda, \mu$ is kept fixed, and only positive integer powers of $\Lambda$ are allowed to vary with cosmic time $t$ and be set equal to $3\, H(t)^2$. This is for covariance reasons explained in \cite{basil,rvm}. 
Doing so, we observe that the so obtained $\rho_\Lambda (H)$, remarkably, adopts precisely the generic RV  structure (\ref{lambda}) around that phase, in which the Ricci scalar is approximated by 
$R\simeq 12H^{2}$, since the Hubble parameter $H$ remains (approximately) constant in this phase.

The imposition of the constraint (\ref{condition}) during the Starobinsky inflationary phase implies, as discussed previously, that the correct phenomenology is attained.
So, if the constraint \emph{was} an \emph{exact result}, the
effective vacuum energy density of the SUGRA model
would correspond to the ${\widehat R}^2 \rightarrow 144 \, H^2$ terms in (\ref{effactionl3})
with (\ref{egc}) playing the r\^ole of the effective gravitational constant,
\begin{equation}\label{rholambda1}
\rho_\Lambda^{\rm SUGRA} (H)^{\rm exact}_{\rm constraint} = \frac{72}{\kappa_{\rm eff}^2} \, \frac{\alpha_2}{\alpha_1} \,  H^4 =
\frac{18}{\kappa_{\rm eff}^2} \, \frac{\alpha_2^F + \alpha_2^B}{\alpha_1^F + \alpha_1^B} \,  H^4~,\end{equation} 
where we used (\ref{alpha}), (\ref{aif}).
The form (\ref{rholambda1}) constitutes an admissible class of RV models (\emph{cf.} Eq.~(\ref{lambda})).
Notice that in Eq.~(\ref{rholambda1}) there is no $\nu$ term. This is important, in the sense that in such a model, as a result of the effective gravitational constant (\ref{egc}) entering the game, which in this scenario~\cite{ahmstaro} is viewed as the `\emph{physica}l'  reduced Planck mass of order $10^{18}$
~GeV, the gravitino  mass and global SUSY breaking scales,
(\ref{scales}), when expressed in terms of $\kappa_{\rm eff}$ are of order one, that is one encounters a Planck-scale gravitino. Despite this, the vanishing of $\nu$ makes the renormalization-group equation (\ref{rholambda1}) a consistent one within the perturbative class of (\ref{lambda}).

However the above construction leads to the absence
of a present-era (small, positive) cosmological constant $c_0$. This arises from the fact that we imposed the constraint
(\ref{condition}) exactly. It may well be that such a condition leaves (non-perturbatively, when all the higher than one-loop contributions are taken into account) a very small (constant in cosmic time) contribution $c_0 > 0$ which is preserved until the present day.  Unfortunately, our one-loop construction does not allow us to explain the magnitude and the sign of this constant term, but this is equivalent to offering a solution to the cosmological constant problem, which of course our approximate one-loop analysis cannot provide. While we do not have a quantitative calculation at this point, nevertheless, the above argument provides at least an interesting qualitative explanation for it: the origin of the current cosmological term $\rho_\Lambda^0$ in the model might well be attributed to quantum (non-perturbative) effects in the SUGRA effective action, which prevent the complete cancellation (\ref{condition}) from being realised. The constant residue $c_0$ is then transferred throughout the cosmic history and pops up in our days in the form of the tiny vacuum energy contribution (\ref{vcc}), as discussed above.

Under this assumption, the initial gravitational coupling $\kappa$, and thus the Einstein term $\frac{1}{2\kappa^2} \int d^4 x \sqrt{-g}\, \widehat R$ would enter the game during the exit phase from inflation~\footnote{The reader should bear in mind that, since during the inflationary phase the scalar degree of freedom of the Starobinsky action is slowly rolling, if there is inflation in the conformally rescaled metric (\ref{confmetric}), there is also inflation in the initial metric. The Starobinsky inflation arguments are also not affected if a small contribution to the cosmological constant, of order of the present-era one, enters the effective action (\ref{steps}), as this is negligible compared to the Hubble scale of inflation.}.
In such a case, in the exit phase, the effective vacuum energy of the SUGRA  model at the inflationary phase should correspond to \emph{both} $\alpha_1$ and $\alpha_2$ terms of (\ref{effactionl3}), with the constraint (\ref{condition}) failing by a tiny amount $\tilde c_0 > 0$ corresponding to the present-era cosmological constant:
\begin{equation}\label{rholambda}
\rho_\Lambda^{\rm SUGRA} (H) = \frac{1}{\kappa^2} \Big( \tilde c_0 + 6 \alpha_1 \, H^2 + 72\, \alpha_2\,  H^4 \Big) \simeq \frac{1}{\kappa^2} \tilde {c}_{0}+ 1.59\, {\tilde \kappa}^{2}
\sigma^{2}_{c}H^{2}+ 12.88\, H^{4} \;,\end{equation}
where we have used the explicit form of the coefficients $\alpha_i$, $i=1,2$ (Eqs.~(\ref{alpha}), (\ref{aif})), with the scales $\mu$ and $\Lambda$  \emph{fixed} through (\ref{scales}). 

Comparing (\ref{rholambda}) with 
(\ref{lambda}) with $c_0= \tilde{c}_0/3$, one obtains  the
effective values for the coefficients $\nu$ and $\alpha$ in the matched-SUGRA/RV model~\cite{basil}:
\be\label{eq:nueff} 
\nu_{\rm eff} \simeq 0.53 \, \kappa^2\tilde{\kappa}^2\sigma_c^2\simeq {\cal O}\left(\frac{m^2_{3/2}}{M^2_{\rm Pl}}\right)\,, \qquad \alpha_{\rm eff} \simeq  4.30\,H_I^2\,\kappa^2\simeq {\cal
O}\left(\frac{H^2_I}{M^2_{\rm Pl}}\right)\,.
\ee
Thus we observe that, within the context of the pure SUGRA model, where only the gravitino plays the r\^ole of ``matter'', both coefficients are small, of typical order $10^{-9}$, in accordance with their interpretation
as $\beta$-function coefficients of the running vacuum energy
density. In the general case where the parameters of the SUGRA model are varied from the generic values considered in (\ref{scales}), but
within the allowed range, the values of $\nu_{\rm eff}$ and
$\alpha_{\rm eff}$ can also undergo some variation and the sign of
$\nu_{\rm eff}$ could change. However we stress that the sign of
$\alpha_{\rm eff}$ remains always positive, which is essential for a
correct description of inflation~\cite{basil}. 

Having matched the evolution of the SUGRA model with the RV flow at the exit of the inflationary Starobinsky phase, one can apply the analysis outlined in the previous subsection \ref{sec:evol}, in order to smoothly connect the inflationary and the current (cosmological-constant dominated) eras of this Universe, with the interpolation of radiation and matter dominated epochs. In this approach, the smallness of the cosmological constant today is attributed to a failure of the constraint (\ref{condition}) due to non-perturbative quantum (supergravity) effects. This offers an interesting new insight into the cosmological constant problem, which, needless to say however, remains unsolved.

\section{Conclusions \label{sec:concl}}

In this talk we considered a minimal inflationary scenario, by means of which a gravitino condensate in supergravity models is held responsible for breaking local supersymmetry dynamically and inducing inflation in an indirect way by means of a Starobinsky-type inflation in the massive gravitino phase. 
The inflaton field in this approach is associated with the scalar mode that collectively parametrizes the effects of the quadratic-curvature contributions to the one-loop quantum effective action of the gravitino condensate, after integrating out graviton and massive gravitino degrees of freedom, in a de Sitter background. 
The model involves parameters that assume values of a natural and phenomenologically relevant order of magnitude, specifically global supersymmetry scale and gravitino masses of the order of GUT mass scales or less. 
Such a scenario is a truly minimal scenario for natural inflation, in the sense that it involves two scalar primordial composite modes to achieve dynamical breaking of a gauge symmetry (supergravity) and inflation. 

Moreover, the effective potential of this model has been cast in a form that allows a running vacuum scenario to be in operation, thereby implying a smooth connection of the inflationary phase to the current era, characterised by a small cosmological constant. The smallness of the latter has been attributed to non perturbative quantum (supergravity) effects that lead to the failure of the constraint (\ref{condition}) characterising the Starobinsky inflationary phase of the model. 

From our analysis it becomes clear that,  in order to ensure phenomenologically relevant supersymmetry breaking scales and gravitino masses, one needs to apply the above ideas to Jordan-frame extensions of the minimal $\mathcal N =1$ $D=4$  SUGRA model, which involve a third scalar field - the dilaton. 
The latter is responsible, through its appropriate potential, for the breaking of the local scale symmetry of the theory. In the context of the next to minimal supersymmetric standard model, which Jordan-frame SUGRA models can incorporate, such dilatons may be composite of appropriate matter superfields, involving Higgs (supermultiplets). 

Details of the microscopic matter model are important in order to ensure the correct cosmological evolution, in particular satisfaction of the Big-Bang-Nucleosynthesis constraints. A GUT scale gravitino can be made to decay fast enough so as not to disturb the BBN, but this depends on the details of the matter sector of the theory. Such topics are left for future investigations. Nevertheless, we believe that the above simple models for dynamical breaking of supergravity  and their links with Starobinsky inflation and Running Vacuum models,
deserve further study, and stand a serious chance of leading to realistic phenomenological scenarios compatible with the cosmological and particle physics data. 

\section*{Acknowledgements}

N.E.M. would like to thank the organisers of the 4th ICNFP 2015 conference in Kolymbari (Crete, Greece)
for their kind invitation to give a plenary talk.
This work is supported in part by the London Centre for Terauniverse Studies (LCTS), using funding from the European Research Council via the Advanced Investigator Grant 267352 and by STFC (UK) under the research grant ST/L000326/1.


\begin{thebibliography}{99}

\bibitem{bau} V.~A.~Kuzmin, V.~A.~Rubakov and M.~E.~Shaposhnikov,
  Phys.\ Lett.\ B {\bf 155}, 36 (1985).
  doi:10.1016/0370-2693(85)91028-7;
  M.~E.~Shaposhnikov,
  Nucl.\ Phys.\ B {\bf 287}, 757 (1987).
  doi:10.1016/0550-3213(87)90127-1;
M.~B.~Gavela, P.~Hernandez, J.~Orloff and O.~Pene,
  Mod.\ Phys.\ Lett.\ A {\bf 9}, 795 (1994)
  doi:10.1142/S0217732394000629
  [hep-ph/9312215].
M.~B.~Gavela, P.~Hernandez, J.~Orloff, O.~Pene and C.~Quimbay,
  Nucl.\ Phys.\ B {\bf 430}, 382 (1994)
  doi:10.1016/0550-3213(94)00410-2
  [hep-ph/9406289].


  
\bibitem{susy} J.~Wess and B.~Zumino,
  Phys.\ Lett.\ B {\bf 49}, 52 (1974).
  doi:10.1016/0370-2693(74)90578-4;
  Nucl.\ Phys.\ B {\bf 70}, 39 (1974).
  doi:10.1016/0550-3213(74)90355-1;

\bibitem{akulov} D.~V.~Volkov and V.~P.~Akulov,
  Phys.\ Lett.\ B {\bf 46}, 109 (1973).
  doi:10.1016/0370-2693(73)90490-5
  

\bibitem{cm} S.~R.~Coleman and J.~Mandula,
  Phys.\ Rev.\  {\bf 159}, 1251 (1967).
  doi:10.1103/PhysRev.159.1251
  
  \bibitem{inflation}  A.~H.~Guth,
  Phys.\ Rev.\ D {\bf 23}, 347 (1981).
  doi:10.1103/PhysRevD.23.347;
  A.~D.~Linde,
  Phys.\ Lett.\ B {\bf 116}, 335 (1982).
  doi:10.1016/0370-2693(82)90293-3;
  \emph{ibid.} {\bf 108}, 389 (1982).
  doi:10.1016/0370-2693(82)91219-9.
 
 
\bibitem{Planck} P.~A.~R.~Ade {\it et al.}  [Planck Collaboration],
  arXiv:1502.02114 [astro-ph.CO].
  
\bibitem{encyclo}  J.~Martin, C.~Ringeval and V.~Vennin,
  arXiv:1303.3787 [astro-ph.CO].
    
    
  \bibitem{natural}   J.~R.~Ellis, D.~V.~Nanopoulos, K.~A.~Olive and K.~Tamvakis,
  Phys.\ Lett.\ B {\bf 118}, 335 (1982).
  
      
    \bibitem{croon} D.~Croon, J.~Ellis and N.~E.~Mavromatos,
  Phys.\ Lett.\ B {\bf 724}, 165 (2013)
  [arXiv:1303.6253 [astro-ph.CO]];
  J.~Ellis, N.~E.~Mavromatos and D.~J.~Mulryne,
  JCAP {\bf 1405}, 012 (2014)
  [arXiv:1401.6078 [astro-ph.CO]].
  
  
  
  

\bibitem{sugra_inflation_review}
  M.~Yamaguchi,
  Class.\ Quant.\ Grav.\  {\bf 28} (2011) 103001
  [arXiv:1101.2488 [astro-ph.CO]].


\bibitem{sugra_hybrid}
  A.~D.~Linde and A.~Riotto,
  Phys.\ Rev.\ D {\bf 56} (1997) 1841
  [hep-ph/9703209],
  E.~Halyo,
  Phys.\ Lett.\ B {\bf 387} (1996) 43
  [hep-ph/9606423].
  
\bibitem{sugra_chaotic}
  M.~Kawasaki, M.~Yamaguchi and T.~Yanagida,
  Phys.\ Rev.\ Lett.\  {\bf 85} (2000) 3572
  [hep-ph/0004243],
  J.~Ellis, M.~A.~G.~Garcia, D.~V.~Nanopoulos and K.~A.~Olive,
  arXiv:1403.7518 [hep-ph],
  R.~Kallosh, A.~Linde and A.~Westphal,
  Phys.\ Rev.\ D {\bf 90}, no. 2, 023534 (2014)
  doi:10.1103/PhysRevD.90.023534
  [arXiv:1405.0270 [hep-th]].
  
  \bibitem{sugra_staro}
  J.~Ellis, D.~V.~Nanopoulos and K.~A.~Olive,
  JCAP {\bf 1310} (2013) 009
  [arXiv:1307.3537],
W.~Buchmuller, V.~Domcke and K.~Kamada,
  Phys.\ Lett.\ B {\bf 726} (2013) 467
  [arXiv:1306.3471 [hep-th]],
  F.~Farakos, A.~Kehagias and A.~Riotto,
  Nucl.\ Phys.\ B {\bf 876} (2013) 187
  [arXiv:1307.1137],
  S.~V.~Ketov and T.~Terada,
  JHEP {\bf 1312} (2013) 040
  [arXiv:1309.7494 [hep-th]];
  JHEP {\bf 1412}, 062 (2014)
  doi:10.1007/JHEP12(2014)062
  [arXiv:1408.6524 [hep-th]].
  C.~Pallis,
  JCAP {\bf 1404} (2014) 024
  [arXiv:1312.3623 [hep-ph]],
  S.~Ferrara, R.~Kallosh and A.~Van Proeyen,
  JHEP {\bf 1311} (2013) 134
  [arXiv:1309.4052 [hep-th]];
 C.~Pallis and N.~Toumbas,
  arXiv:1512.05657 [hep-ph].

    
  \bibitem{confsugra} S.~Ferrara, R.~Kallosh, A.~Linde, A.~Marrani and A.~Van Proeyen,
  Phys.\ Rev.\ D {\bf 82}, 045003 (2010)
  [arXiv:1004.0712 [hep-th]];
  Phys.\ Rev.\ D {\bf 83} (2011) 025008
  [arXiv:1008.2942 [hep-th]].

\bibitem{sugrainfl} 
M.~A.~G.~Garcia and K.~A.~Olive,
  JCAP {\bf 1309}, 007 (2013)
  doi:10.1088/1475-7516/2013/09/007
  [arXiv:1306.6119 [hep-ph]];
  D.~Roest, M.~Scalisi and I.~Zavala,  
  JCAP {\bf 1311}, 007 (2013)
  doi:10.1088/1475-7516/2013/11/007
  [arXiv:1307.4343];
S.~Ferrara, R.~Kallosh, A.~Linde and M.~Porrati,
  S.~Ferrara, R.~Kallosh, A.~Linde and M.~Porrati,
  Phys.\ Rev.\ D {\bf 88}, no. 8, 085038 (2013)
  doi:10.1103/PhysRevD.88.085038
  [arXiv:1307.7696 [hep-th]]; 
  JCAP {\bf 1311}, 046 (2013)
  doi:10.1088/1475-7516/2013/11/046
  [arXiv:1309.1085 [hep-th]].
  
\bibitem{staro} 
  A.~A.~Starobinsky,
  Phys.\ Lett.\ B {\bf 91}, 99 (1980).


 \bibitem{Freedman}
  D.~Z.~Freedman and A.~Van Proeyen,
  \emph{Supergravity} 
  (Cambridge, UK: Cambridge Univ. Pr. (2012)).

\bibitem{Nieuwenhuizen}
  P.~Van Nieuwenhuizen,
  Phys.\ Rept.\  {\bf 68} (1981) 189.
  
   
  \bibitem{emdyno} 
J.~Ellis and N.~E.~Mavromatos,
  Phys.\ Rev.\ D {\bf 88}, 085029 (2013)
  [arXiv:1308.1906 [hep-th]].

  

\bibitem{ahm} J.~Alexandre, N.~Houston and N.~E.~Mavromatos,
  Phys.\ Rev.\ D {\bf 88}, 125017 (2013)
  [arXiv:1310.4122 [hep-th]]; 
  Int.\ J.\ Mod.\ Phys.\ D {\bf 24}, no. 04, 1541004 (2015)
  doi:10.1142/S0218271815410047
  [arXiv:1409.3183 [gr-qc]].
   

\bibitem{ahmstaro} 
  J.~Alexandre, N.~Houston and N.~E.~Mavromatos,
  Phys.\ Rev.\ D {\bf 89}, 027703 (2014)
  [arXiv:1312.5197 [gr-qc]].

\bibitem{terada} S.~V.~Ketov and T.~Terada,
  JHEP {\bf 1307}, 127 (2013)
  doi:10.1007/JHEP07(2013)127
  [arXiv:1304.4319 [hep-th]];
  K.~Hamaguchi, T.~Moroi and T.~Terada,
  Phys.\ Lett.\ B {\bf 733}, 305 (2014)
  doi:10.1016/j.physletb.2014.05.006
  [arXiv:1403.7521 [hep-ph]].
  

 \bibitem{fradkin}
  E.~S.~Fradkin and A.~A.~Tseytlin,
  Nucl.\ Phys.\ B {\bf 234} (1984) 509.


\bibitem{basil} S.~Basilakos, N.~E.~Mavromatos and J.~Sol\`a,
  arXiv:1505.04434 [gr-qc].
 
 
\bibitem{rvm}  I. L. Shapiro and J. Sol\`{a}, {Phys. Lett}.
    {\bf B530}, 10 (2002) [arXiv:hep-ph/0104182];  
 Phys.\ Lett.\ B {\bf 682}, 105 (2009)
  [arXiv:0910.4925; \emph{ibid.} {\bf B475}, 236 (2000) [hep-ph/9910462];  
  JHEP {\bf 0202} 006 (2002) [hep-th/0012227]; 
  J.~Sol{\`a},  J. Phys. Math. Theor. {\bf A41}, 164066  (2008) [arXiv:0710.4151]; 
 I.~L. Shapiro, J. Sol\`a, H.
\v{S}tefan\v{c}i\'{c} JCAP 0501 012 (2005);
See also: J. Sol\`a,
J. Phys. Conf. Ser. {\bf 453} (2013)  012015 [arXiv:1306.1527];   
J. Sol\`a, and A. G\'omez-Valent,
Int. J. of Mod. Phys. {\bf D24} (2015) 1541003
 [e-Print: arXiv:1501.03832];
 J.~A.~S.~Lima, S.~Basilakos and J.~Sol\`a,
  Mon.\ Not.\ Roy.\ Astron.\ Soc.\  {\bf 431}, 923 (2013)
  [arXiv:1209.2802];
  S.~Basilakos, J.~A.~S.~Lima and J.~Sol\`a,
  Int.\ J.\ Mod.\ Phys.\ D {\bf 22}, 1342008 (2013)
  [arXiv:1307.6251]; \emph{ibid.}  {\bf D23} (2014) 1442011 [arXiv:1406.2201].


\bibitem{Wetterich}
  J.~Jaeckel and C.~Wetterich,
  Phys.\ Rev.\ D {\bf 68} (2003) 025020
  [hep-ph/0207094].
  

\bibitem{NJL}
Y.~Nambu and G.~Jona-Lasinio,
  Phys.\ Rev.\  {\bf 122} (1961) 345.
  
   
\bibitem{DeserZumino}
  S.~Deser and B.~Zumino,
  Phys.\ Rev.\ Lett.\  {\bf 38} (1977) 1433.




\bibitem{Komargodski}
  Z.~Komargodski and N.~Seiberg,
  JHEP {\bf 0909} (2009) 066
  [arXiv:0907.2441 [hep-th]].



\bibitem{nmssm} M.~B.~Einhorn and D.~R.~T.~Jones,
  JHEP {\bf 1003}, 026 (2010)
  [arXiv:0912.2718 [hep-ph]].






\bibitem{higgsinfl} F.~L.~Bezrukov and M.~Shaposhnikov,
  Phys.\ Lett.\ B {\bf 659}, 703 (2008)
  [arXiv:0710.3755 [hep-th]];
  F.~Bezrukov, A.~Magnin, M.~Shaposhnikov and S.~Sibiryakov,
  JHEP {\bf 1101}, 016 (2011)
  [arXiv:1008.5157 [hep-ph]].



\bibitem{odintsov}
  I.~L.~Buchbinder and S.~D.~Odintsov,
  Class.\ Quant.\ Grav.\  {\bf 6} (1989) 1955;
see also  S.~D.~Odintsov,
  Phys.\ Lett.\ B {\bf 213}, 7 (1988).



\bibitem{aben}
  I.~Antoniadis, C.~Bachas, J.~R.~Ellis and D.~V.~Nanopoulos,
  Phys.\ Lett.\ B {\bf 211} (1988) 393.
  



\bibitem{loop}   P.~C.~W.~Davies, S.~A.~Fulling, S.~M.~Christensen and T.~S.~Bunch,
  Annals Phys.\  {\bf 109}, 108 (1977);
  T.~S.~Bunch and P.~C.~W.~Davies,
  Proc.\ Roy.\ Soc.\ Lond.\ A {\bf 360}, 117 (1978);
 N. Birrell and P. Davies, \emph{Quantum Fields in Curved Space} 
(Cambridge Monogr. Math. Phys., 1982).



\bibitem{whitt} K.~S.~Stelle,
  Gen.\ Rel.\ Grav.\  {\bf 9}, 353 (1978);
  B.~Whitt,
  Phys.\ Lett.\ B {\bf 145}, 176 (1984).



\bibitem{witten} E.~Witten,
  Nucl.\ Phys.\ B {\bf 188}, 513 (1981).
 
 


  
  \end{thebibliography}
  \end{document}